\documentclass[aps,prl,amsfonts,twocolumn,preprintnumbers,%
               superscriptaddress,nofootinbib]{revtex4}

\usepackage{times}

\newcommand{\be}{\begin{equation}}
\newcommand{\ee}{\end{equation}}

\newcommand{\ba}{\begin{eqnarray}}
\newcommand{\ea}{\end{eqnarray}}

\begin{document}

\preprint{SLAC-PUB-10565}
\preprint{WM-04-112}
\preprint{UMN-D-04-5}

\title{Constraints on proton structure
from precision atomic physics measurements}

\author{Stanley J. Brodsky}
\email[]{sjbth@slac.stanford.edu}
\affiliation{Stanford Linear Accelerator Center, Stanford University,
Stanford, CA 94309}

\author{Carl E. Carlson}
\email[]{carlson@physics.wm.edu}
\affiliation{Particle Theory Group, Physics Department,
College of William and Mary, Williamsburg, VA 23187-8795}

\author{John R. Hiller}
\email[]{jhiller@d.umn.edu}
\affiliation{Department of Physics, University of Minnesota-Duluth,
Duluth, MN 55812}

\author{Dae Sung Hwang}
\email[]{dshwang@sejong.ac.kr}
\affiliation{Department of Physics, Sejong University, Seoul 143-747,
Korea}

\date{July 22, 2005 (revised)}

\begin{abstract}

Ground-state hyperfine splittings in hydrogen and muonium are very well
measured. Their difference, after correcting for magnetic moment and
reduced mass effects, is due solely to proton structure---the large QED 
contributions for a pointlike nucleus essentially cancel.
The rescaled hyperfine difference depends on the Zemach radius, a
fundamental measure of the proton, computed as an integral over a
product of electric and magnetic proton form factors. The
determination of the Zemach radius, $(1.043 \pm 0.016) {\rm\ fm}$, from
atomic physics tightly constrains fits to accelerator measurements of
proton form factors. Conversely, we can use muonium data to extract an
``experimental'' value for QED corrections to hydrogenic hyperfine
data; we find that measurement and theory are consistent.

\end{abstract}


\maketitle


{\it Introduction.} Quantum Electrodynamics, QED, stands
out as the most precisely tested component of the Standard Model.
QED predictions for the classic Lamb shift, and hyperfine splittings (hfs)
in hydrogen, positronium, and muonium have been confirmed to better
than 10 parts per million (ppm)~\cite{Karshenboim:2003jr,Sapirstein:gn},
2 ppm~\cite{Sapirstein:gn,Faustov:yp},
2 ppm~\cite{Karshenboim:2003jr,Sapirstein:gn},
and 1 part in 10 million~\cite{Karshenboim:2003jr}, respectively.  The
measurements of the electron and positron gyromagnetic ratios agree
with order-$\alpha^4$ perturbative QED predictions to 1 part in
$10^{11}$~\cite{Kinoshita1990}.   QED and gauge theory have thus been
validated to extraordinary precision.

In this paper we shall show how one can combine precision atomic physics
measurements to determine a fundamental property of the proton to remarkable
precision. The difference between the ground-state hfs of
hydrogen and muonium, after correcting for the different magnetic moments
of the muon and the proton and for reduced mass effects, is due to the
structure of the proton. The QED contributions for a pointlike
nucleus essentially cancel. The largest proton structure contribution
to the hfs difference is proportional to the Zemach
radius~\cite{Zemach,Friar:2003zg}, which can be computed as an
integral over the product of the elastic electric and magnetic form
factors of the proton. The remaining proton structure corrections,
the polarization
contribution~\cite{Faustov:yp,Iddings,Drell:1966kk,DeRafael:mc,Gnaedig:qt}
from inelastic states in the spin-dependent virtual Compton amplitude and
the proton size dependence of the relativistic recoil
corrections~\cite{Bodwin:1987mj,Bodwin:1984dr}, have small
uncertainties. As we shall show, the resulting high precision determination
of the Zemach radius from the atomic physics measurements provides an
important constraint on fits to accelerator measurements of the proton
electric and magnetic form factors.

An important motivation for examining form factor constraints comes from
the recent polarization transfer measurements of the proton electric form
factor $G_E(Q^2)$~\cite{Jones:1999rz,Gayou:2001qt,Gayou:2001qd}. The
polarization transfer results are at variance with the published
Rosenbluth measurements of $G_E$. The difference may well be due to
corrections from hard two-photon exchange~\cite{Chen:2004tw,other2photon}.
One wants to examine with the maximum possible precision whether the new
determinations of $G_E(Q^2)$, falling with respect to $G_M(Q^2)$,
is compatible with other information on the form factor. 
The extraction
of the Zemach radius to be described here provides such a constraint.


{\it A sum rule for proton structure.} We now show how one can use the
hfs of the muonium atom $(e^- \mu^+)$ to expose the hadronic structure
contributions to the hydrogen hfs.  For an electron bound to a positively
charged particle of mass $m_N$, magnetic moment $\mu_N=(g_N/2) (e/2m_N)$,
and Land\'e $g$-factor $g_N$, the leading term in the hfs is the Fermi energy,
\be \label{eq:Fermi}
E_F^N=\frac{8}{3\pi}\alpha^3 \mu_B\mu_N \frac{m_e^3 m_N^3}{(m_N+m_e)^3}.
\ee
Here, ``$N$'' stands for either the $p$ or $\mu^+$ nucleus.  By convention,
the exact magnetic moment $\mu_N$ is used for the proton or muon, but only
the lowest order term, the Bohr magneton $\mu_B$, is inserted for the $e^-$.

The ground-state hydrogen hfs can be written as
\ba \label{eq:Ephfs}
E_{\rm hfs}(e^-p) &=&
    (1+\Delta_{\rm QED}+\Delta_R^p+\Delta_S \\
    && +\Delta_{hvp}^p+\Delta_{\mu vp}^p+\Delta_{\rm weak}^p)E_F^p,
    \nonumber
\ea
where $\Delta_{\rm QED}$ represents QED corrections, $\Delta_R^p$
represents recoil effects, including finite-size recoil corrections,
$\Delta_S$ represents the proton structure contributions,
and $\Delta_{hvp}^p$, $\Delta_{\mu vp}^p$, and $\Delta_{\rm weak}^p$
represent the effects of hadronic vacuum polarization, muonic
vacuum polarization, and weak interactions, respectively.
The corresponding quantity for muonium is simply
\be \label{eq:Emuhfs}
E_{\rm hfs}(e^-\mu^+)=(1+\Delta_{\rm QED}+\Delta_R^\mu+\Delta_{hvp}^\mu
   +\Delta_{\rm weak}^\mu)E_F^\mu .
\ee

We define the fractional difference between the hydrogen and rescaled
muonium hfs as
\ba \label{eq:DeltaF}
\Delta_{\rm hfs}
   &\equiv& \frac{E_{\rm hfs}(e^-p)}{E_{\rm hfs}(e^-\mu^+)}
         \frac{\mu_\mu}{\mu_p}\frac{(1+m_e/m_p)^3}{(1+m_e/m_\mu)^3}-1
                                \\
    &=&\frac{E_{\rm hfs}(e^-p)/E_F^p}{E_{\rm hfs}(e^-\mu^+)/E_F^\mu}-1 .
    \nonumber
\ea
The large contributions from QED  corrections cancel in $\Delta_{\rm hfs}$.
Since the hfs of hydrogen and muonium, as well as the
ratio of muon and proton magnetic moments, have been measured to better
than 30 ppb, $\Delta_{\rm hfs}$ can be determined to high precision from
experiment.

From Eqs.~(\ref{eq:Ephfs}) and (\ref{eq:Emuhfs}), we have
\ba
\lefteqn{\frac{E_{\rm hfs}(e^-p)/E_F^p}{E_{\rm hfs}(e^-\mu^+)/E_F^\mu}}&& \\
 && =
\frac{(1+\Delta_{\rm QED}+\Delta_R^p+\Delta_S+\Delta_{hvp}^p
   +\Delta_{\mu vp}^p+\Delta_{\rm weak}^p) }
  {(1+\Delta_{\rm QED}+\Delta_R^\mu+\Delta_{hvp}^\mu
   +\Delta_{\rm weak}^\mu) }  .  \nonumber
\ea
Thus we can obtain a result for the proton structure contribution in terms 
of quantities measurable to high precision in atomic physics:
\ba \label{eq:constraint}
\Delta_S&=&\Delta_{\rm hfs} 
    + \Delta_R^\mu +\Delta_{hvp}^\mu+\Delta_{\rm weak}^\mu \\
    &&- \left(\Delta_R^p +\Delta_{hvp}^p+\Delta_{\mu vp}^p
                          +\Delta_{\rm weak}^p\right) \nonumber \\
    &&+\Delta_{\rm hfs} \left( \Delta_{\rm QED} + \Delta_R^\mu
                   +\Delta_{hvp}^\mu+\Delta_{\rm weak}^\mu \right) .
                   \nonumber
\ea
The cross terms are smaller than the uncertainties in the leading terms,
and here $\Delta_{\rm QED}$ can be approximated as $\alpha/2\pi$.

The proton structure contributions consist of the classic Zemach term
computed from a convolution of elastic form factors and the polarization
contribution from the inelastic hadronic states contributing to the
spin-dependent virtual Compton scattering:
$\Delta_S =  \Delta_Z + \Delta_{\rm pol}$.
In addition, as we discuss below, the relativistic recoil corrections
of order $\alpha m_e / m_p$ are modified by the finite size of the
proton. The Zemach term takes into account the finite-size correction
to the proton magnetic interactions as well as the finite-size
distortions of the electron's orbit in the hydrogen
atom~\cite{Zemach,Friar:2003zg}:
$\Delta_Z=-2\alpha m_e \langle r \rangle_Z
    \left( 1 + \delta_Z^{\rm rad} \right)$,
where $\langle r \rangle_Z$ is the radius of the proton as calculated from the
Zemach integral
\be
\langle r \rangle_Z
    = - \frac{4}{\pi}\int_0^\infty\frac{dQ}{Q^2}
    \left[  G_E(Q^2) \frac{G_M(Q^2)}{1+\kappa_p} -1\right],
\label{Zemachint}
\ee
with $G_E$ and $G_M$ the electric and magnetic form factors of the proton, 
normalized with $G_E(0) = G_M(0)/(1+\kappa_p) = 1$, and $\kappa_p=g_p/2-1$.  
Additionally, $\delta_Z^{\rm rad}$ is a radiative correction to the Zemach 
term estimated in~\cite{Bodwin:1987mj}. It has been calculated analytically 
in~\cite{Karshenboim:1996ew} for the case where the form factors are 
represented by dipole forms:
 $\delta_Z^{\rm rad}
= (\alpha/3\pi)
    \left[ 2 \ln ( \Lambda^2/ m_e^2 ) - 4111/420 \right]$.
With $\Lambda^2 = 0.71$ GeV$^2$, this yields
$\delta_Z^{\rm rad} = 0.0153$.

The main part of the inelastic contribution can be constructed from
the work of Iddings~\cite{Iddings} and Drell and Sullivan~\cite{Drell:1966kk}.
Compact expressions are given by De Rafael~\cite{DeRafael:mc},
Gn\"adig and Kuti~\cite{Gnaedig:qt}, and
Faustov and Martynenko~\cite{Faustov:yp}
in terms of the Pauli form factor $F_2$ and spin-dependent structure functions
$g_1$ and $g_2$ of the proton.


\begin{table*}

\caption{Proton electric charge radius $\sqrt{\langle r^2_E\rangle}$,
Zemach contribution $\Delta_Z$ to the hfs, and Zemach radius
$\langle r \rangle_Z$ for various
parameterizations of
$G_E$ and $G_M$.  The results should be compared to
$\Delta_Z = - (40.0\pm 0.6)$ ppm or
$\langle r \rangle_Z = (1.043 \pm 0.016)$ fm, as obtained from
analysis of atomic hfs data.
The dipole form is $G_M(Q^2)=(1+\kappa_p)/(1+Q^2/0.71\,{\rm GeV}^2)^2$.
The $G_E$ labeled JLab is~\cite{Gayou:2001qt}
$\left(1-0.13\frac{Q^2}{{\rm GeV}^2}\right)\frac{G_M}{1+\kappa_p}$.
Parameterizations A-I and A-II are from \cite{Arrington:2003qk}.
Those labeled Brodsky-Carlson-Hiller-Hwang (BCHH), I and II, use
$F_2/F_1=[1/\kappa_p^2+Q^2/(1.25\,{\rm GeV})^2]^{-1/2}$
and $F_2/F_1=\kappa_p[1+(Q^2/0.791\,{\rm GeV}^2)^2\ln^{7.1}(1+Q^2/4m_\pi^2)]
/[1+(Q^2/0.380\,{\rm GeV}^2)^3\ln^{5.1}(1+Q^2/4m_\pi^2)]$,
respectively \cite{Brodsky:2003gs}.
The last column gives the contribution to $\langle r \rangle_Z$ from $Q>0.8$ GeV.}

\label{tab:Zemach}

\begin{ruledtabular}

\begin{tabular}{lccccc}

\multicolumn{2}{l}{Parameterizations} & $\sqrt{\langle r^2_E\rangle}$ 
                                                            & $\Delta_Z$
  & \multicolumn{2}{l}{$\langle r \rangle_Z$ (fm)} \\
$G_M$ & $G_E$ &  \ \ (fm)\ \ &\ \ (ppm) \ \  &\ \ total \ \   
                                                       &  \ $Q>0.8$ GeV   \\
\hline

dipole\ & $G_M/(1+\kappa_p)$\ & 0.811 & --39.32  &   1.025  &  0.310 \\
dipole  & JLab                & 0.830 & --39.83  &   1.038  &  0.310 \\
A-I     & A-I                 & 0.868 & --41.46  &   1.081  &  0.310 \\
A-I     & $G_M/(1+\kappa_p)$  & 0.863 & --41.34  &   1.077  &  0.310 \\
A-II    & A-II                & 0.829 & --40.29  &   1.050  &  0.310 \\
A-II    & JLab                & 0.855 & --40.70  &   1.061  &  0.310 \\
dipole  & BCHH-I              & 0.789 & --38.85  &   1.012  &  0.310 \\
A-II    & BCHH-I              & 0.816 & --39.72  &   1.035  &  0.309 \\
dipole  & BCHH-II             & 0.881 & --40.91  &   1.066  &  0.310 \\
A-II    & BCHH-II             & 0.905 & --41.77  &   1.088  &  0.310 \\
\end{tabular}

\end{ruledtabular}

\end{table*}




{\it Evaluation of the constraint.} We will consider each term on the
right hand side of Eq.~(\ref{eq:constraint}).
To compute $\Delta_{\rm hfs}$ from (\ref{eq:DeltaF}), we use
the measured hydrogen hfs~\cite{Karshenboim:1997zu}
$E_{\rm hfs}(e^-p) = 1~420.405~751~766~7(9) {\rm\ MHz}$
and muonium hfs~\cite{Liu:1999iz}
$E_{\rm hfs}(e^- \mu^+) = 4~463.302~765(53) {\rm\ MHz}$.
The measured masses are~\cite{Eidelman:2004wy}
$ m_p = 938.272~029(80) {\rm\ MeV}$,
$m_\mu = 105.658~369(9) {\rm\ MeV}$, and
$m_e = 0.510~998~918(44) {\rm\ MeV}$.
The ratio of magnetic moments has been measured to high precision,
$\pm 0.028$ ppm; the value obtained without input from the
muonium hfs is~\cite{MohrPC}
$\mu_\mu/\mu_p = 3.183~345~20(20)$.
From these values we find $\Delta_{\rm hfs}=145.51(4)$ ppm.

The recoil corrections $\Delta_R^N$ are separated into 
relativistic corrections $\Delta_{\rm rel}^N$ and
additional radiative corrections $\Delta_{\rm rad}^N$.
The order-$\alpha$ relativistic recoil correction $\Delta_{\rm rel}^N$ has
been computed by Arnowitt~\cite{Arnowitt} for muonium $(N=\mu)$.
Bodwin and Yennie~\cite{Bodwin:1987mj} quote the corrections
to second order in $\alpha$ in their Eq.~(1.10), which is analogous
to Eq.~(\ref{eq:drp}) below.  Expressions
for the radiative correction $\Delta_{\rm rad}^\mu$ are given
in \cite{Kinoshita} and \cite{Eides}.
With use of~\cite{Eidelman:2004wy} $\alpha^{-1} = 137.035~999~11(46)$
and~\cite{Bennett:2004pv} $\kappa_\mu = 0.001~165~920~8(6)$, 
the total correction is evaluated to be $\Delta_R^\mu=-178.34$ ppm.

Bodwin and Yennie~\cite{Bodwin:1987mj} have also computed the corrections to 
their formula in the hydrogen case due to the finite size of the proton from 
elastic intermediate states.  Note that these are finite-size corrections to 
the recoil correction and are distinct from the Zemach correction.  A mark of 
the distinction is that after scaling out the lowest order Fermi hfs, the 
recoil corrections go to zero as $(m_p/m_e) \to \infty$, whereas the Zemach 
correction does not. The Bodwin--Yennie pointlike result to order $\alpha^2$ 
is~\cite{Bodwin:1987mj}
\ba        \label{eq:drp}
\Delta_{\rm rel}^p &=& {\alpha\over \pi}{m_e m_p \over m^2_p - m^2_e}
  \left(-3 + 3 \kappa_p -{9\over 4} {\kappa^2_p\over 1+\kappa_p} \right)
  \ln \frac{m_p}{m_e}
                    \nonumber \\
&&+~ \alpha^2 \frac{m_e}{m_p} \Bigg\{
   2 \ln { 1\over 2\alpha } - 6 \ln{2} + \frac{65}{18}
                    \nonumber \\
&&+~
 \kappa_p \left[
   \frac{7}{4} \ln { 1\over 2\alpha } - \ln{2} + \frac{31}{36}
 \right] \nonumber \\
&& +~ {\kappa_p \over 1+\kappa_p} \left[
 -\frac{7}{4} \ln { 1\over 2\alpha } + 4 \ln{2} - \frac{31}{8}
 \right] \Bigg\},
\ea
with~\cite{Eidelman:2004wy} $\kappa_p = 1.792~847~351(28)$.  This gives 
$\Delta_{\rm rel}^p = (-2.01 + 0.46) {\rm\ ppm}$, where the two terms are from 
${\cal O}(\alpha)$ and ${\cal O}(\alpha^2)$.  Quoting~\cite{Bodwin:1987mj}, 
finite-size corrections change this to 
$\Delta_{\rm rel}^p = (+ 5.22(1) + 0.46) {\rm\ ppm} = 5.68(1) {\rm\ ppm}$, 
where the quoted error is an estimate using the dipole form factor for the
proton (both $G_E$ and $G_M$) with mass parameter 
$\Lambda^2 = 0.71 \pm 0.02$ GeV$^2$. 
An additional radiative correction~\cite{Karshenboim:1996ew} of 0.09 ppm 
brings $\Delta_R^p$ to 5.77 ppm.

Volotka {\it et al.}~\cite{Volotka:2004zu} have reevaluated the finite-size
corrections to the proton recoil corrections with the same magnetic radius, 
but with a charge radius taken from Ref.~\cite{Sick:2003gm}, and find 
$\Delta_R^p = 5.86$ ppm, or 0.18 ppm larger than Bodwin and Yennie. 
By forcing the magnetic form factor to reproduce their result for
the Zemach integral, Volotka {\it et al.} obtain a second value of
6.01 ppm.  We shall use the first Volotka result and
include an uncertainty of $0.15$ ppm to 
cover the difference between the modified Bodwin--Yennie
and the second Volotka determinations.  
Note that structure-dependence uncertainty within the recoil
corrections is still well under the uncertainty of the polarization terms, 
and that this uncertainty in the recoil term can be reduced as knowledge of 
the form factors improves.

Estimates of the weak and vacuum polarization corrections are also given
by Volotka {\it et al.}~\cite{Volotka:2004zu}.  From these and 
from the individual values for $\Delta_{\rm hfs}$, $\Delta_R^\mu$, and
$\Delta_R^p$, we obtain $\Delta_S = -38.62(16)$ ppm.  
Thus the contribution of proton structure is constrained by atomic
physics with an uncertainty well under one percent.


{\it The Zemach term.} We shall subtract the polarization contributions to 
isolate the Zemach term and then explore its relevance to new form factor 
parameterizations. Although the polarization contributions have been long 
known to be small~\cite{DeRafael:mc,Gnaedig:qt}, the error in $\Delta_Z$ is 
essentially all due to the uncertainty in $\Delta_{\rm pol}$.  From Faustov 
and Martynenko~\cite{Faustov:yp}, we take 
$\Delta_{\rm pol} = 1.4 {\rm\ ppm\ } \pm 0.6 {\rm\ ppm}$, which implies 
$\Delta_Z =- (40.0 \pm 0.6) {\rm\ ppm}$ and thus 
$\langle r \rangle_Z = (1.043 \pm 0.016) {\rm\ fm}$. 
The unit conversion used $\hbar c = 197.326~968 (17)$ MeV-fm.

Predictions for $\Delta_Z$ and $\langle r \rangle_Z$ as computed from a
selection of parameterizations of the form factors are given in
Table~\ref{tab:Zemach}. The first row is the textbook standard, wherein
both $G_M$ and $G_E$ are given by the dipole form. The result,
$\Delta_Z = -39.32$ ppm $=-38.72 (1+\delta_Z^{\rm rad})$ ppm, 
can already be found in \cite{Bodwin:1987mj}.
New analytic fits to the form
factors~\cite{Arrington:2003qk,Brodsky:2003gs} make a significant
change in the Zemach integral, of up to 6\%. 
The form factor
parameterization given in \cite{Sick:2003gm} yields~\cite{Friar:2003zg}
$\langle r \rangle_Z = 1.086(12)$ fm.  
It is not clear why the large difference exists.  The scattering data is 
subject to radiative and other corrections; any difference highlights the 
usefulness of having the precise value that we have derived.
Not all of the $\Delta_Z$ or
$\langle r \rangle_Z$ for the different models in the table are
compatible with the results extracted from the analysis of the atomic data.   
However, the $G_M$-$G_E$ combination suggested in the third row from 
the end of the Table 
shows that fully compatible models exist.

The table also shows results for the charge radius
$\sqrt{\langle r^2_E\rangle}=\sqrt{-6 {d \over dQ^2}G_E(Q^2)\vert_{Q^2=0}}$.
The values compare to results from Lamb shift
measurements~\cite{Pachucki} (0.871(12) fm), a continued-fraction
fit to $G_E$~\cite{Sick:2003gm} (0.895(18) fm), a standard empirical
fit~\cite{Simon:1980hu} (0.862(12) fm),  
and the 2002 Committee on Data for Science and Technology 
value~\cite{Mohr:2000ie} (0.8750(68) fm).

The differences among the Zemach integrals for different form factors
derive mainly from the lower $Q$ range of the integral, where the
different parameterizations of $G_E$ are less variant, but their effect
on the integral is greater.  This is seen in the last two columns of
Table~\ref{tab:Zemach}.  About 30\% of the Zemach integral comes from
$Q$ above 0.8 GeV, but little of this has to do with the form factors. 
Recall that the numerator of the Zemach integrand is
$G_E\,G_M/(1+\kappa_p)- 1$, and for high $Q$ the form factors fall
away, leaving the ``$-1$.'' In the region above 0.8 GeV, the ``$-1$''
contributes 0.314 fm. 

Two fits by Arrington~\cite{Arrington:2003qk} are used in the Table,
denoted A-I and A-II.  Fit A-I uses only Rosenbluth data and
A-II uses $G_E/G_M$ from the polarization
results~\cite{Jones:1999rz,Gayou:2001qt,Gayou:2001qd}.  While A-II
represents the data well overall, for $Q$ below 0.8 GeV its
$G_E/G_M$ ratio falls too quickly by nearly a factor of two compared
to the actual polarization data. The same is true for the fit
denoted JLab~\cite{Gayou:2001qt}.


{\it Discussion.} In this paper we have shown how one can combine
high-precision atomic physics measurements of the ground-state hydrogen
and muonium hfs and the ratio of muon to proton magnetic moments to
isolate the proton structure contributions. In our method, the
theoretically complex QED corrections to the bound-state problem do not
appear~\cite{atomic_ex}. Remarkably, the total proton structure contribution
$\Delta_S =-38.62(16)$ ppm  to the hydrogen hfs is determined to better than
$1\%.$ Since the polarization contribution can be determined from the
measured spin-dependent proton structure functions  $g_1(x,Q^2)$ and
$g_2(x,Q^2)$, we obtain a precise value for the Zemach radius
$\langle r \rangle_Z = (1.043 \pm 0.016)$ fm, which is defined from a
convolution of the $G_E$ and $G_M$ form factors. This
new determination gives an important constraint on the analytic form
and fits to the proton form factors at small $Q^2$.  The precision of
the Zemach radius will be further improved when new, more precise data
for $g_1$ and $g_2$, especially at small $\nu$ and $Q^2$, becomes
available.

The proton structure terms can also be extracted using the hydrogen hfs
alone~\cite{Volotka:2004zu,dupays}.  The Zemach radius obtained this
way is slightly smaller but consistent with our result.

Conversely, by combining the muonium and hydrogen hfs data,
one can obtain an ``experimental''
value for the purely QED bound-state radiative corrections:
$\Delta_{\rm QED} = 1 136.09(14) {\rm\ ppm}$.
To minimize the uncertainty, we take advantage of the 
measured ratio~\cite{Mohr:2000ie} $m_p/m_e=1836.152~672~61(85)$.
This value of $\Delta_{\rm QED}$ is consistent with
the calculated QED correction used in~\cite{Volotka:2004zu,dupays}. 

Our method of combining experimental atomic physics has other
applications; for example, measurements of the difference of the Lamb
shifts (or Rydberg spectra) of muonium and hydrogen could potentially
give a very precise value for the proton's electric charge radius,
since again the QED radiative corrections essentially cancel.
Similarly, the difference of lepton anomalous moments $a_\mu - a_e$
directly exposes the hadronic and weak corrections to the muon moment.

\acknowledgments
This work was supported in part by the U.S. Department of Energy
contracts DE-AC02-76SF00515 (S.J.B.), 
and
DE-FG02-98ER41087 (J.R.H.); by the U.S. National Science Foundation
Grant PHY-0245056 (C.E.C); and by the KISTEP (D.S.H.).  We thank
John Arrington, Todd Averett, Geoffrey Bodwin, Michael Eides, 
Lee Roberts, Ralph Segel, Barry Taylor, Marc Vanderhaeghen,
and Andrey Volotka for helpful remarks.


\end{document}